\documentclass[prl,twocolumn,showpacs,preprintnumbers,superscriptaddress,
amsmath,amssymb,tightenlines]{revtex4}

\usepackage{bm}
\usepackage{graphicx}

\newcommand{\rv}{{\bf r}}

\newcommand{\beq}{\begin{equation}}
\newcommand{\eeq}{\end{equation}}
\newcommand{\bea}{\begin{eqnarray}}
\newcommand{\eea}{\end{eqnarray}}

\newcommand{\<}{\langle}
\renewcommand{\>}{\rangle}

\newcommand{\commentout}[1]{{}}

\begin{document}

\title{Dirac monopoles and dipoles in ferromagnetic
spinor Bose-Einstein condensates}

\author{C. M. Savage}
\affiliation{Australian Centre for Quantum Atom Optics, Australian National University, ACT 0200, Australia}
\email{craig.savage@anu.edu.au}
\author{J. Ruostekoski}
\affiliation{Department of Physical Sciences, University of
Hertfordshire, Hatfield, Herts, AL10 9AB, UK}
\email{j.ruostekoski@herts.ac.uk}

\begin{abstract}
We investigate a radial spin hedgehog, analogous to the Dirac monopole, in an optically trapped atomic spin-1 Bose-Einstein condensate. By joining together a monopole-antimonopole pair, we may form a vortex line with free ends. We numerically simulate the three-dimensional dynamics and imaginary time relaxation of these structures to nonsingular textures and show they can be observable for realistic experimental parameters.
\end{abstract}

\pacs{03.75.Lm,03.75.Mn}

\date{\today}
\maketitle

\section{Introduction}

Topologically interesting structures in quantum fields range from vortices and vortex lattices in single component fields \cite{ANG02}, to Skyrmions in multi-component fields \cite{SAV03}. They have long been important in the study of superfluid physics \cite{Vollhardt} and quantum cosmology \cite{Preskill}. Experimental dilute gas Bose-Einstein condensates (BECs) allow these structures to be investigated with unprecedented flexibility.

\begin{figure}[!b]
\includegraphics[width=\columnwidth]{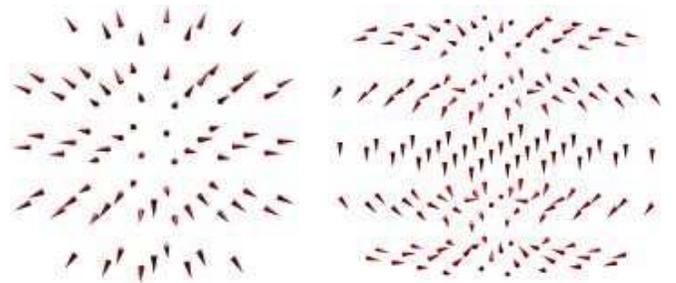}
\caption{Spin fields $\langle  {\bf F} \rangle$ for the monopole of Eq.~(\ref{monopole}) (left) and the dipole of Eq.~(\ref{dipole}) (right). In all figures the positive $z$ axis is upward, and cones are proportional to $\langle  {\bf F} \rangle$. The dipole's singular points at $ c = \pm  (2.5/\sqrt{2}) x_{ho}$ are near the second and fourth planes of spin vectors.}
\label{initial spins}
\end{figure}

The successful trapping of atomic BECs by purely optical means has opened up a fascinating domain of research  \cite{STA98,Kuppens}. Unlike in magnetic traps, where the spin of the atoms is effectively
frozen, in an optical dipole trap the magnetic degrees of freedom dramatically affect the structure of the condensates. In particular, their equilibrium states exhibit richer degeneracy of physically distinguishable states than magnetically trapped BECs, with the degeneracy, or order, parameter depending on the atomic spin. In this paper we study a ferromagnetic spin-1 BEC in $^{87}$Rb \cite{Kuppens} whose degeneracy parameter is determined by the rotations of the spin \cite{Ho,Ohmi}.  We show that in the classical mean-field, or Gross-Pitaevskii, approximation there can exist a vortex line with a free end, terminating on a hedgehog-like spin configuration whose superfluid velocity profile coincides with the vector potential of the Dirac magnetic monopole \cite{Dirac,Felsager,Blaha}. Magnetic monopoles are important in quantum cosmology; explaining their observed low density stimulated the theory of cosmological inflation \cite{Guth}. The terminating vortex line, known as the Dirac string, is possible in a ferromagnetic spinor BEC because the noncommutativity of the spatial rotations results in ambiguity of the condensate phase difference between two spatial points. We also show that a monopole can be attached to an anti-monopole with a vortex line, such that neither end reaches the boundary of the atomic cloud, forming a ``dipole'', Fig.~\ref{initial spins}.

In a magnetically trapped BEC the order parameter, determined by the complex scalar field, has a well-defined phase. Hence, its circulation around any closed loop can only equal an integer multiple of $2\pi$, where the integer coefficient represents a topologically invariant winding number. As a result, any vortex line is either closed or terminates at the boundary of the atomic cloud. In a ferromagnetic spinor BEC the order parameter symmetry group is determined by the spatial rotations of the spin to be SO(3). This can be represented by a local coordinate axis, or a {\it triad}. A vortex line in a ferromagnetic BEC is determined by an integer multiple of $2\pi$ spatial rotations of the triad about the direction of the spin, when one moves around any closed loop surrounding the vortex core. Due to the noncommutativity of the spatial rotations, such a vortex is not a topological invariant, but can be continuously deformed, e.g., to a vortex line with the opposite sign or to a disgyration. Since the relative condensate phase between two spatial locations is not uniquely defined, a vortex line may also terminate in the middle of the atomic cloud. This is analogous to the possibility of vortex lines with free ends in the angular momentum texture of superfluid liquid $^3$He-A \cite{VOL76}.

In this paper we consider the mean-field model of a ferromagnetic spin-1 BEC. We numerically integrate the Gross-Pitaevskii equations in real time to find the dynamics, and in imaginary time to find energetic minima. Our initial states are based on analytic expressions for the monopoles and dipoles, with overall Thomas-Fermi density profiles. In $^3$He studies dipoles have been referred to as monopolium \cite{Soni}. In that context it was suggested that monopolium lattices might be dynamically stable in rotating systems.

Previous theoretical work on monopoles in atomic BECs has been limited to anti-ferromagnetic systems \cite{Stoof,Chang, RUO03}, two-component monopoles \cite{Martikainen}, or analogous two-dimensional structures called ``spin monopoles'' \cite{BUS99,Garcia-Ripoll}. This work has shown that monopoles may be created using phase imprinting techniques for creating solitons and vortices \cite{Chang, Martikainen,RuostekoskiAnglin,Ruostekoski}. Spin textures have also been recently experimentally created in an atomic spin-1 $^{23}$Na BEC by means of rotating the spin profile with spatially dependent external magnetic fields~\cite{LEA02}.

\section{Mean-field model for monopoles and dipoles}

The second quantized Hamiltonian for the spin-1 BEC in an optical dipole trap reads \cite {Ho,Ohmi,Pethick,Robins,Law}
\begin{multline} \label{hamiltonian}
 H = \int  d^3 r \left\{
-\frac{\hbar^2}{2m} \nabla \hat{\psi}_\alpha^\dagger \cdot \nabla \hat{\psi}_\alpha
+ V({\bf r}) \hat{\psi}_\alpha^\dagger \hat{\psi}_\alpha
\right.  \\
+ g_F \mu_B \hat{\psi}_\alpha^\dagger {\bf B} \cdot {\bf F}_{\alpha \beta} \hat{ \psi}_{\beta}
+ \frac{1}{2} c_0
\hat{\psi}_\alpha^\dagger \hat{\psi}^\dagger_{\beta} \hat{ \psi}_{\beta} \hat{\psi}_{\alpha} \\
\left.  + \frac{1}{2} c_2
\hat{\psi}_\alpha^\dagger \hat{\psi}^\dagger_{\beta} {\bf F}_{\alpha \gamma} \cdot  {\bf F}_{\beta \delta} \hat{\psi}_{\delta} \hat{\psi}_{\gamma} \right\} \, ,
\end{multline}
where $\hat{\psi}_\alpha$ is the field annihilation operator for the $\alpha$ Zeeman component and $m$ is the atomic mass. $V({\bf r})$ is the optical trapping potential. In experiments, optical traps with a wide range of aspect ratios and depths have been demonstrated \cite{Gorlitz}. In this paper we consider an isotropic harmonic trapping potential with frequency $\omega$: $V({\bf r})=m\omega^2 r^2/2$. The magnetic field vector is denoted by ${\bf B}$, $\mu_B$ is the Bohr magneton, and the Land\'{e} factor $g_F = -1/2$ for the $F=1$ transition of $^{87}$Rb. In Eq.~(\ref{hamiltonian}), ${\bf F}$ is the vector formed by the three components of the $3\times3$ Pauli spin-1 matrices \cite {Pethick}, and we have used the Einstein summation convention over repeated indices. The Zeeman energy is correct to first order in the magnetic field. $c_0$ and $c_2$ are the spin-independent and spin-dependent two-body interaction coefficients. In terms of the s-wave scattering lengths $a_0$ and $a_2$ for the channels with total angular momentum zero and two they are: $c_0 = 4\pi\hbar^2(a_0 +2a_2)/3m$ and $c_2 = 4\pi\hbar^2(a_2 -a_0)/3m$. We use the values given by van Kempen {\it et al.}  \cite{vanKempen} for $^{87}$Rb: $a_0= 101.8 a_B$ and $a_2= 100.4 a_B$, so that $(a_0 +2a_2)/3 = 100.9 a_B$ and $(a_2 -a_0)/3 = -0.47 a_B$, where the Bohr radius $a_B = 0.0529$ nm.
We also use $\omega = 2 \pi\times 10$ s$^{-1}$, and assume a total number of $N=10^6$ atoms. We make our lengths dimensionless with the length scale $x_{ho} = \sqrt{\hbar/m\omega}$. As usual, the dynamics is invariant up to changes of length and time scale under changes of $\omega$ and $N$ provided $N^2 \omega$ is constant.

In the mean-field Gross-Pitaevskii approximation the order parameter is the three-component spinor $\Psi(\rv)=\sqrt{n(\rv)}\zeta(\rv)$, with $\zeta(\rv)^\dagger \zeta(\rv)=1$, and $n(\rv)$ is the total density. Then the expectation value of the spin is given by $\< {\bf F} \>=\zeta_\alpha^* {\bf F}_{\alpha\beta} \zeta_\beta$. The contribution of the two-body interaction to the energy is
\beq \label{two body energy}
 E_{2b} = \int  d^3 r \left\{
\frac{1}{2} n^2 ( c_0 + c_2 \< {\bf F} \>^2 ) \right\} \, .
\eeq
The preceding parameters for $^{87}$Rb correspond to the ferromagnetic case, since $c_2 < 0$, and the energy is minimized when $| \langle {\bf F} \rangle |=1$ throughout the BEC for the case of a uniform spin distribution.

For the ferromagnetic state $| \langle  {\bf F} \rangle | = 1$, all the degenerate, but physically distinguishable, states are related by spatial rotations of the atomic spin. These may be conveniently defined by means of the three Euler angles~\cite{Ho}:
\begin{align}  \label{ferromagnetic spinor}
\zeta &=
\begin{pmatrix}
\zeta_1 \\ \zeta_0 \\ \zeta_{-1}
\end{pmatrix}
=  e^{i\varphi} e^{-i F_z \alpha} e^{-i F_y \beta} e^{-i F_z \tau}
\begin{pmatrix}
1 \\ 0 \\ 0
\end{pmatrix}
\notag \\
&= e^{i\varphi'}
\begin{pmatrix}
e^{-i\alpha}\cos^2{(\beta/2)} \\
\sin{(\beta)}/\sqrt{2} \\
 e^{i\alpha}\sin^2{(\beta/2)}
\end{pmatrix}
\end{align}
where $\varphi' \equiv \varphi-\tau$ and $\varphi$ denotes the macroscopic condensate phase. The combination $\varphi-\tau$ indicates an invariance under the change of the macroscopic phase if it is simultaneously accompanied by a spin rotation $\tau$ of the same magnitude. The equivalence of the phase change and the spin rotation represents a broken relative gauge-spin symmetry. Consequently, the symmetry group of the order parameter is fully determined by the Euler angles to be SO(3). The direction of the expectation value of the spin is $\<{\bf F}\>= \hat{z} \cos{\beta}+\sin{\beta}(
\hat{x} \cos{\alpha} +\hat{y} \sin{\alpha} )$, where $\hat{x}, \hat{y}, \hat{z}$ are the Cartesian coordinate unit vectors.

\begin{figure}
\includegraphics[width=\columnwidth]{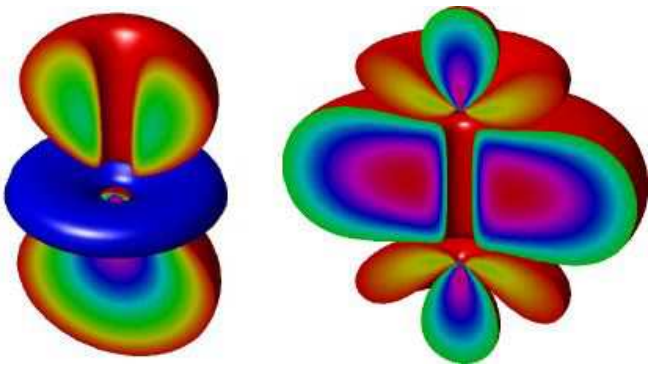}
\caption{Spinor densities for the monopole (left) and dipole (right). The spinors are given by Eq.~(\ref{monopole}) and Eq.~(\ref{dipole}) respectively, multiplied by the Thomas-Fermi and vortex density profiles, as described in the text. All lobes are bounded by density isosurfaces for a spinor component for $x < 0$. On the $x=0$ plane the isosurfaces are capped with a density colormap for the corresponding spinor component. Monopole: the top lobe is for $ | \Psi_1 |^2$. The density zero associated with the doubly quantized vortex is seen along the positive $z$ axis.  The bottom lobe is for $ | \Psi_{-1} |^2$, and the central lobe is for $ | \Psi_0 |^2$. The singly quantized vortex can be seen. Dipole: the top and bottom lobes are for $ | \Psi_1 |^2$, the central one for $ | \Psi_{-1} |^2$, and the remaining two lobes for $ | \Psi_0 |^2$. The vortices in $\Psi_{-1}$ and $ \Psi_0$ can be seen on the $z$-axis. }
\label{isosurfaces}
\end{figure}

We are interested in the ferromagnetic state with a singular pointlike core in the atomic spin field. Using the radial, azimuthal, and polar spherical coordinates $(r, \theta, \phi)$, we select $\beta=\theta$, $\alpha=\phi$, and $\varphi'=-\phi$ in Eq.~(\ref{ferromagnetic spinor}), yielding
\beq \label{monopole}
\zeta=
\begin{pmatrix}
e^{-2i\phi}\cos^2{(\theta/2)}\\
 e^{-i\phi} \sin{(\theta)}/\sqrt{2}\\
 \sin^2{(\theta/2)}
\end{pmatrix}  \, .
\eeq
The corresponding spin exhibits a radially outward hedgehog field and forms the monopole $\<{\bf F}\>= \hat{z} \cos{\theta}+\sin{\theta}( \hat{x} \cos{\phi} +\hat{y} \sin{\phi} )$. However,
for $\theta=0$ the value of $\phi$ is not defined and the first component is singular. Therefore, the point singularity, or the hedgehog, at $r = 0$ is attached to a line singularity (in this case a doubly-quantized SO(3) vortex) along the positive $z$ axis. This is called a Dirac string.
Note that for $z<0$ there are no line singularities. The structure of the monopole in terms of the individual spin components is: a doubly-quantized vortex line in the $\Psi_1$ component, a vortex with unit circulation in the $\Psi_0$ component, and no vortex in the $\Psi_{-1}$ component. The corresponding spinor densities are shown in Fig.~\ref{isosurfaces}.

The corresponding superfluid velocity is
\begin{align}
\< {\bf v}_s \> &= -i{\frac{\hbar}{M} }\zeta^\dagger \nabla\zeta
={\frac{\hbar}{M}} (\nabla \varphi' -\cos{\beta}\nabla\alpha)  \notag \\
&=-\frac{\hbar(1+\cos{\theta})} {Mr\sin{\theta}} \,\hat{\phi} \, ,
\end{align}
where $\hat{\phi}$ is the polar unit vector. This has the form of the electromagnetic vector potential of the Dirac magnetic monopole, analogous to the corresponding angular momentum texture in superfluid liquid helium-3 \cite{VOL76}.

An anti-monopole with spin vectors pointing radially inward $\<{\bf F}\>= -\hat{z} \cos{\theta}-\sin{\theta}( \hat{x} \cos{\phi} +\hat{y} \sin{\phi} )$, and the Dirac string along the negative $z$ axis, is given by
\beq \label{anti-monopole}
\zeta=
\begin{pmatrix}
e^{-2i\phi}\sin^2{(\theta/2)}\\
-e^{-i\phi} \sin{(\theta)}/\sqrt{2}\\
 \cos^2{(\theta/2)}
\end{pmatrix} \, .
\eeq

An interesting system results from attaching a monopole to an antimonopole
by joining their Dirac strings. The resulting vortex line with two free ends then no longer reaches the boundary of the atomic
gas.  From the spin field given by Soni \cite{Soni} follows an expression for the dipole spinor
\beq \label{dipole}
\zeta= \frac{1}{2}
\begin{pmatrix}
1+(r^2 \cos 2 \theta -c^2)/\lambda \\
e^{i\phi} \sqrt{2} ( r^2 \sin 2 \theta )/  \lambda \\
e^{2i\phi} \{1-(r^2 \cos 2 \theta -c^2)/\lambda \}
\end{pmatrix} \, ,
\eeq
where
\beq
\lambda = \sqrt{(r^2 -c^2) +4r^2 c^2 \sin^2 \theta} \, .
\eeq
This has monopole and antimonopole type singularities on the $z$-axis at $z=|c|$ and $z= -|c|$ respectively. On the $z$-axis for $|z| > |c|$, $\zeta_1 = 1$ and $\zeta_{-1} = 0$, while between the poles, i.e. for $|z| < |c|$, $\zeta_1 = 0$, and $\zeta_{-1}$ is singular, representing a doubly-quantized SO(3) vortex line. The corresponding spinor densities are shown in Fig.~\ref{isosurfaces}.

The topological stability of linear and pointlike topological defects may be characterized by the first and second homotopy groups. Since the second homotopy group of SO(3) is trivial, there are no stable topological point defects. However, as we previously showed, we may form the monopole by attaching it to a singular vortex line. It is not possible to form an isolated point defect in SO(3). The `charge' $W$ of the Dirac monopole is similar to the topological invariant of the point defects, which is determined by the way the order parameter behaves on a closed surface enclosing the defect \cite{Vollhardt}:
\beq\label{W}
W= {1\over 8\pi} \int d \Omega_i\,\epsilon_{ijk}
\langle  {\bf F} \rangle
 \cdot  \left( {\partial{\langle  {\bf F} \rangle}\over\partial x_j} \times
 {\partial{\langle  {\bf F} \rangle}\over\partial x_k} \right) \,.
\eeq
Here $\epsilon_{ijk}$ denotes the totally antisymmetric tensor. 
The integration is defined over a closed surface enclosing the monopole. The singular vortex line always cuts through the surface. However, as we show in the following section, the Dirac monopole can decay to a nonsingular coreless texture in which case $\langle {\bf F} \rangle$ is well-defined everywhere.

\section{Numerical Results}

We now turn to the numerical studies of the classical mean-field theory for the ferromagnetic spinor BEC. We integrate the relaxation and dynamics of the monopole and the monopole-antimonopole pair.
The Gross-Pitaevskii equations for the dynamics of the spinor mean-field are \cite {Pethick,Robins}
\begin{multline} \label{GP}
i \, \hbar \frac{\partial \Psi_\alpha}{\partial t} =
-\frac{\hbar^2}{2m} \nabla^2 \Psi_\alpha + V({\bf r}) \Psi_\alpha
+g_F \mu_B {\bf B} \cdot {\bf F}_{\alpha \beta} \Psi_{\beta}
 \\
+ c_0 \Psi^\dagger_{\beta} \Psi_{\beta} \Psi_{\alpha} +
c_2 \Psi^\dagger_{\beta} {\bf F}_{\beta \gamma} \cdot  {\bf F}_{\alpha \delta} \Psi_{\delta} \Psi_{\gamma} \, .
\end{multline}
The explicit forms of the two-body interaction terms, for $\alpha = 1,0,-1$ are, respectively,
\begin{gather}
( c_0 + c_2) n \Psi_1 -2 c_2 | \Psi_{-1} |^2 \Psi_1  + c_2 \Psi_{-1}^{*} \Psi_0^2 \, , \notag \\
( c_0 + c_2) n \Psi_{0} -c_2 | \Psi_{0} |^2  \Psi_{0}  + 2 c_2 \Psi_{0}^{*} \Psi_1 \Psi_{-1}  \, , \notag \\
( c_0 + c_2) n \Psi_{-1} -2 c_2 | \Psi_{1} |^2 \Psi_{-1}  +c_2 \Psi_{1}^{*} \Psi_0^2 \, .
\end{gather}
We solve these equations numerically in three spatial dimensions starting from initial monopoles and dipoles to find the dynamical evolution of the system. We also solve them in imaginary time $\tilde{t} = -it$ to find minima, either local or global, of the system's energy functional. This requires normalization of the Gross-Pitaevskii spinor after each time step.  Our numerical method is pseudo-spectral, using FFTs to evaluate the Laplacian, and a forth-order Runge-Kutta time step. It is described in detail in Ref.~\cite{Otago}. All our simulations were performed using $128^3$ spatial grids for each of the three spinor components. The physical region simulated was centered on the trap, and was typically $(25/\sqrt{2}) x_{ho}$ long in each dimension. A typical time step was $10^{-3}/\omega$.

The initial conditions for our numerical integrations of the Gross-Pitaevskii equations (\ref{GP}) are either the monopole Eq.~(\ref{monopole}) or dipole  Eq.~(\ref{dipole}) with an overall Thomas-Fermi density profile $n( {\bf r} )$, multiplied by an approximation to the vortex density profiles \cite{Pethick}, see Fig.~\ref{isosurfaces}. In both real and imaginary time the structures initially evolve by the filling in of the vortex core on the $z$ axis by the vortex-free component: for the monopole this is the $\Psi_{-1}$ component, and for the dipole the $\Psi_{1}$ component.

\begin{figure}
\includegraphics[width=\columnwidth]{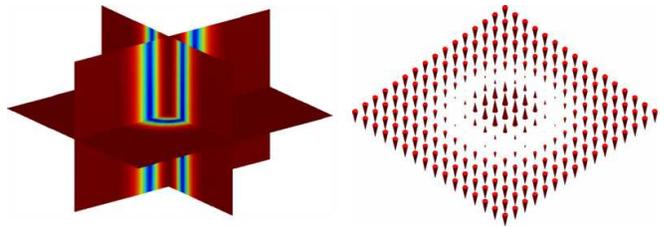}
\caption{Spin magnitude $| \langle  {\bf F} \rangle |$ and spin field $\langle  {\bf F} \rangle$ of the relaxed state, with a magnetic field $ | {\bf B} |  =  \sqrt{2} \hbar \omega / \mu_B$, of either an initial monopole (${\bf B}$ in $\hat{z}$ direction)  or dipole (${\bf B}$ in $-\hat{z}$ direction). The imaginary time simulations were run for $-it = 30/\omega$, with a time step $0.002/\omega$. Other parameters are as given in the text. Left. 3D slice plot of $| \langle  {\bf F} \rangle |$ in the central region. The width shown in each dimension is half of the total simulation width of $(25/\sqrt{2}) x_{ho}$. The slice planes are $x=0$, $y=0$, and $z=0$. The colormap corresponds to $| \langle {\bf F} \rangle |$ ranging between 0.046 and 1. Right. $\langle  {\bf F} \rangle$, for the dipole, on the plane $z=0$. Quarter the simulation width is shown in the $x$ and $y$ dimensions. }
\label{spin magnitudes}
\end{figure}

\begin{figure}
\includegraphics[width=\columnwidth]{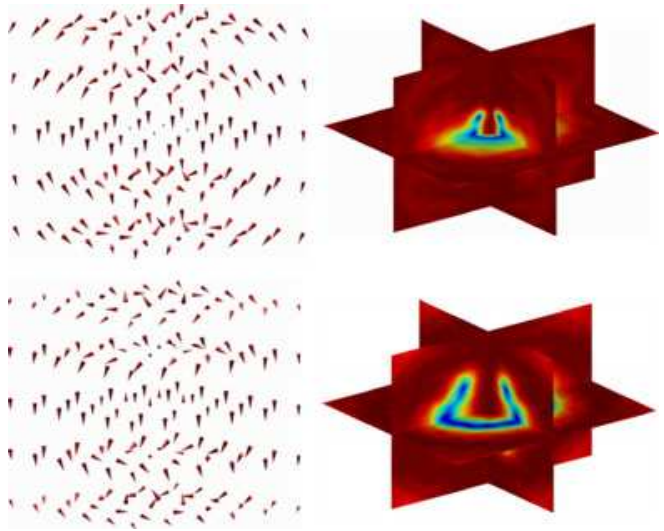}
\caption{Spin fields $\langle  {\bf F} \rangle$ (left column) and spin magnitudes $| \langle  {\bf F} \rangle |$ (right column) after real time evolution of the dipole of Fig.~\ref{initial spins} for $\omega t =$ : 6 (top row), and 12 (bottom row). The spin magnitudes are shown on slice planes $x=0$, $y=0$, and $z=0$. The colormap corresponds to $| \langle  {\bf F} \rangle |$ ranging between 0 and 1. At $\omega t = 6$ the minimum of  $| \langle  {\bf F} \rangle |$ is 0.0033. At $\omega t = 12$ it is $0.0019$. Approximately the central three-quarters of each simulation dimension is shown. Parameters given in text.}
\label{dipole spin}
\end{figure}

The imaginary time evolution relaxes towards the lowest energy component. In the absence of a magnetic field this is the vortex-free component; $\Psi_{-1}$ for the monopole. This represents the relaxation of the monopole configuration towards a uniform spin texture. A magnetic field along the $z$ axis can increase the energy of the vortex-free spin component, reversing the direction of the relaxed spin. However, the vortex cores may still be filled by the vortex-free component $\Psi_{-1}$. This results in a coreless vortex, similar to the Anderson-Toulouse  (AT) spin texture in superfluid liquid helium-3 \cite{AND77,Vollhardt,Mizushima}. However, the system no longer remained in the ferromagnetic ground state and regions of reduced spin magnitude, close to $| \<{\bf F}\> |=0$, were found in both the imaginary and real time simulations, see Figs.~\ref{spin magnitudes} and \ref{dipole spin}.

The relaxation of the singular monopole to a nonsingular coreless texture may seem surprising at first, but it is topologically allowed. We can see how the singular monopole (\ref{monopole}) may be deformed to a nonsingular spin texture by continuously rotating the direction of the spin at the positive $z$ axis. Consider the following parametrization:
\beq \label{monopole2}
\zeta=
\begin{pmatrix}
e^{-2i\phi}\cos^2{[\beta(\rho,z,s)/2]}\\
 e^{-i\phi} \sin{[\beta(\rho,z,s)]}/\sqrt{2}\\
 \sin^2{[ \beta(\rho,z,s)/2 ]}
\end{pmatrix}\,,
\eeq
where $(\rho,\phi,z)$ are the cylindrical coordinates. We write for $z\geq 0$: $\beta=\arctan{(\rho/z)}+s\pi\exp{(-C\rho/z)}$, where $C$ is here chosen to be much larger than the inverse of the radius of the atomic cloud. We continuously evolve the parameter $0\leq s \leq 1$. At $s=0$, we have the Dirac monopole with the singularity at $\rho=0$, $z\geq 0$. The singularity can be removed by continuously evolving $s$ from 0 to 1. At $s=1$, the spin has rotated by the value of $\pi$ at $\rho=0$, $z\geq 0$, and the singularity has disappeared.

Depending on the functional dependence of the parameter $\beta$ we may have different coreless textures. If $\beta(\rho)$ depends on $\rho$ alone, with $\beta(0)=\pi$ and $\beta=0$ at the boundary of the atomic cloud, we have the AT coreless vortex. On the other hand, with $\beta=\pi/2$ at the boundary, we obtain the Mermin-Ho texture \cite{MER76}. For the coreless textures, $W$, in Eq.~(\ref{W}), would represent a topologically invariant winding number, provided that $\<{\bf F}\>$ was asymptotically restricted on a plane. However, unlike in superfluid liquid helium-3, where the interactions between the superfluid and the walls of the container may fix the boundary values of the spin, in atomic BECs the spin may generally rotate and change the value of $W$.

The continuous deformation of the singular Dirac monopole to a nonsingular texture is related to the topology of line defects in ferromagnetic spin-1 BECs. The ground state manifold, with the symmetry group SO(3), corresponds topologically to a three-sphere (a sphere in four dimensions) with the diametrically opposite points set to be equal, $S^3/Z_2$. In this space the only closed paths which cannot be continuously deformed to a point are precisely the paths joining the diametrically opposite points of the three-sphere \cite{comment}. Hence, only vortices with winding number equal to one are topologically stable \cite{Ho} and, e.g., a vortex line with winding number equal to two can be continuously deformed to a vortex-free state.

The regions of low spin in the AT vortex represent the largest values of the spin gradient in the texture, indicating a strong order parameter bending energy. The reduced spin value results from the very weak ferromagnetic energy of the $^{87}$Rb spin-1 condensate: it is energetically more favorable for the system to violate the ferromagnetic spin constraint than to create a spin texture with a very large spin gradient. The regions with a large variation in the spin orientation continuously mix the ferromagnetic $| \<{\bf F}\> |=1$ and the polar $| \<{\bf F}\> |=0$ phases of the BEC, see Fig.~\ref{spin magnitudes}. While the violation of the spinor phase and the variation of the spin value has been previously investigated in the case of singular defects \cite{RUO03}, it is interesting to observe that this may happen even as a result of the order parameter bending energy of nonsingular textures.

In the presence of cylindrically asymmetric perturbations we found the coreless AT texture to be unstable with respect to drifting towards the boundary of the atomic cloud. This is analogous to the slow drift of a vortex line in a single-component BEC in a nonrotating trap \cite{ANG02}. We did not perform any quantitative studies to determine whether the drift of the spin texture could possibly be suppressed by the external magnetic fields.

The dipole relaxes to a state which is the same as in the case of the monopole, apart from reversed spin directions due to the reversed magnetic field.
Although the monopole and dipole relax to the same type of state, their imaginary time evolution follows different paths. In particular the cylindrical shell of low spin magnitude develops differently. For the monopole the transition from $\<{\bf F}\>= -\hat{z}$ in the core to $\<{\bf F}\>= \hat{z}$ at the boundary occurs by the spin vectors in the interfacial region rotating radially outward. For the dipole, however, the direction of spin rotation differs in the regions  $z > 0$ and  $z < 0$. The spin rotates radially outward for $z > 0$, and radially inward for $z < 0$. To accommodate these different rotation directions, near  $z=0$ the spin magnitude drops close to zero early in the relaxation. The cylindrical shell of low spin magnitude then grows in the $ \pm z$ directions as the relaxation progresses.  In contrast, for the monopole the growth occurs from the gas boundaries towards the $z=0$ plane.

%

\begin{figure}
\includegraphics[width=\columnwidth]{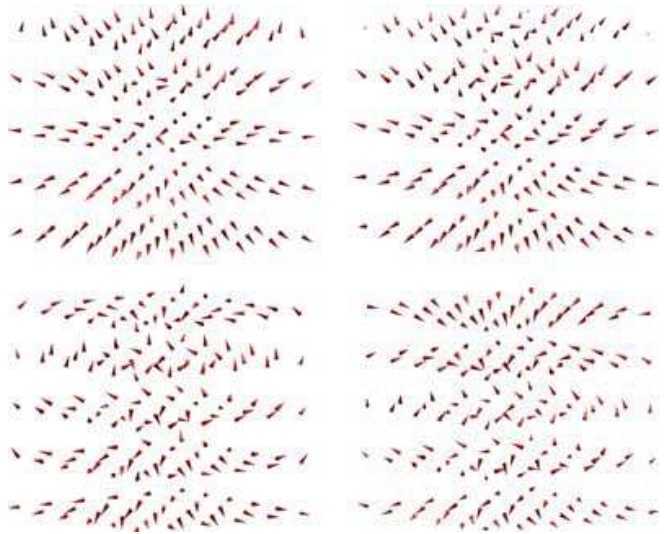}
\caption{Spin fields $\langle  {\bf F} \rangle$ after real time evolution of the monopole of Fig.~\ref{initial spins} for $\omega t =$ : 3 (top left), 6 (top right), 12 (bottom left), 18 (bottom right). Parameters given in text.}
\label{monopole spin}
\end{figure}

The real time evolution of the dipole and monopole, without any applied magnetic field, are shown in Figs. \ref{dipole spin} and \ref{monopole spin}. The initial states were first relaxed for an imaginary time $i\tilde{t} = 0.1/\omega$ before the real time evolution. This ensured that the total density was close to the steady state distribution, hence suppressing density oscillations. The dynamics is quite complex and hence we restrict ourselves to describing the major features. The spin field of the dipole is generally more robust in the dynamical evolution than that of the monopole. Apart from the filling of the vortex core by the vortex-free component, recognizable structures survive around a half period ($t=\pi/\omega$) of evolution. However, the original states are noticeably different after about a period of evolution, which for our parameters is $2\pi / \omega = 0.1$ s. This is likely to be an upper bound on the time available to experiment with these structures. At $\omega t = 18$, approximately three harmonic oscillator periods, the spin field of the initial monopole has developed an approximate dipolar structure, Fig.~\ref{monopole spin}. However the spinor is completely different to the dipole of Eq.~(\ref{dipole}). A major difference between the dynamics of the monopole and dipole is that the dipole develops large regions centered around the $z=0$ plane where the spin magnitude $| \langle {\bf F} \rangle |$ drops to zero, see Fig. \ref{dipole spin}.

We have examined the relaxation and dynamics of Dirac monopoles
and dipoles in trapped ferromagnetic spinor BECs, for realistic
parameters of $^{87}$Rb. There are several interesting aspects to
be explored further. We have not investigated the dynamical
stability of the monopole-antimonopole pair or the effects of
rotation. As proposed by Soni \cite{Soni}, analogous structures,
although not energetically stable in the presence of dissipation,
could still be dynamically stable in rotating superfluid liquid
helium-3. In the absence of dissipation, our dynamical evolution
also suggests the dipole spin field to be surprisingly robust.
However, as our studies have shown, due to the weak ferromagnetic
energy of $^{87}$Rb atom, any dynamically stable dipole
configuration would also include regions of significantly reduced
spin magnitude, presenting a unique and highly nontrivial
topological object for further studies.

\acknowledgments
{This research was supported by the EPSRC and the Australian Partnership for Advanced Computing. ACQAO is an  Australian Research Council Centre of Excellence.}


\begin{thebibliography}{99}
\bibitem{comment} This corresponds to the fact that the fundamental homotopy group of SO(3) has two elements $\pi_1[SO(3)]=Z_2$.
\end{thebibliography}
\end{document}